\documentclass[sigconf,screen]{acmart}

\renewcommand\footnotetextcopyrightpermission[1]{}
\acmConference[CHI '26 TfT Workshop]{CHI '26 Tools for Thought Workshop}{April 2026}{Barcelona, Spain}

\usepackage{booktabs}
\usepackage{graphicx}
\usepackage{caption}
\title{Material for Thought: Generative AI as an Active Creative Medium}

\author{Hugo Andersson}
\affiliation{
  \institution{Aarhus University}
  \city{Aarhus}
  \country{Denmark}
}
\email{hugo@cs.au.dk}

\author{Niklas Elmqvist}
\affiliation{
  \institution{Aarhus University}
  \city{Aarhus}
  \country{Denmark}
}
\email{elm@cs.au.dk}

\begin{document}

%% Abstract & Keywords

\begin{abstract}
    Human-AI collaboration research has largely positioned the human as a judge of AI output, centering effort on evaluating whether recommendations are reliable enough to accept. This decision-support framing leaves little room for the human as creator.
    We argue that for creative work, this framing misdirects human effort toward evaluating correctness rather than exploring and shaping the creative space.
    Drawing on Sch\"{o}n's theory of reflective practice, we propose an alternative: treating generative AI as an active creative medium. As a potter works with clay, humans Shape, Observe, Stir, and Select (\textsc{SOSS}) their medium through ongoing conversation.
    Where generative AI actively tends toward convergence and resolution, the human role of disruption and curation becomes essential for sustaining creative quality.
    We present a creative writing probe, \textsc{Loom}, in which users orchestrate simulated narrative agents. We also introduce the \textsc{SOSS} framework for this mode of engagement, and discuss design implications.
\end{abstract}

\begin{CCSXML}
<ccs2012>
   <concept>
       <concept_id>10003120.10003121.10003125.10011752</concept_id>
       <concept_desc>Human-centered computing~Interaction design theory, concepts and paradigms</concept_desc>
       <concept_significance>500</concept_significance>
   </concept>
</ccs2012>
\end{CCSXML}

\ccsdesc[500]{Human-centered computing~Interaction design theory, concepts and paradigms}
% \keywords{human-AI collaboration, generative AI, tools for thought, creative writing, metacognition, orchestration}

\maketitle

%% ============================================================
%% INTRO
%% ============================================================
\section{Introduction}

What if generative AI is not a tool for thought, but the material for it?

Large language models are no longer emerging technologies; they are routine tools and have become widespread in creative workflows~\cite{mckinsey2025stateofai}. 
The question is not whether writers, designers, and thinkers will work with generative AI, but whether the workflows we design around it will preserve or erode human agency, or at least keep a balance between the two.
This makes the design of human-AI creative workflows a problem of today, not a speculative one. 

A potter does not command clay.
They center it, press into it, feel how it resists and yields, and shape it through an ongoing exchange of intention and response.
The clay has properties---weight, moisture, grain---that constrain and afford.
It is not inert, but it has no agenda of its own.
The creative act lives in the shaping: reading the material, recognizing possibility, knowing when to press and when to let the wheel carry.
Sch\"{o}n, in his theory of reflective practice~\cite{schon1983}, regards this as a conversation with the materials of the situation:
the practitioner makes a move, the material talks back, and the practitioner responds to what they see and feel.
The skill is not in planning the perfect action but in the interactive and emergent process between artist and medium.

However, the prevailing approach to human-AI collaboration frames AI as an advisor and the human as a judge of its output, engaged in \emph{trust calibration}~\cite{zhang2020effect, vaccaro2024}: is this recommendation reliable enough to act on?
This framing has driven extensive research into explainability, confidence communication, and decision support interfaces, all aimed at helping humans better assess when to rely on AI.
Yet evidence suggests this approach is struggling.
Bansal et al.\ found that AI explanations did not improve performance over simply displaying confidence; more troubling, explanations increased human agreement with AI recommendations regardless of their correctness~\cite{bansal2021}, suggesting that richer evaluative information may deepen rather than correct the problem.
Buçinca et al.\ showed that cognitive forcing functions can reduce but not eliminate overreliance on AI, and that the most effective interventions were also the least preferred~\cite{bucinca2021}.
Zhang et al.\ found that while confidence scores can calibrate trust, even successful calibration failed to improve joint decision outcomes when human and AI error boundaries were aligned~\cite{zhang2020effect}.
A recent meta-analysis consolidates these findings at scale: Vaccaro et al.\ synthesized 370 effect sizes across 106 experiments and found that human-AI combinations perform significantly worse than the best of human or AI alone~\cite{vaccaro2024}.

However, Vaccaro et al.'s analysis also revealed a striking asymmetry.
\emph{Decision tasks}, where humans chose among options after receiving AI input, showed significant performance losses. 
\emph{Creation tasks}, on the other hand, where humans generated open-ended content with AI, showed performance gains. 
The difference was statistically significant, yet creation tasks represented only about 10\% of the studied configurations.
Given this asymmetry, how should we design human-AI collaboration around shaping possibility rather than evaluating output?

We propose \emph{AI as an active creative medium}: systems in which AI does not advise a human decision-maker but simulates a possibility space that humans curate.
Our contributions are:
(1)~a theoretical framing of generative AI as active creative medium, drawing on Sch\"{o}n's reflective practice to reposition the human role from evaluator to orchestrator;
(2)~the SOSS framework (Shape, Observe, Stir, Select) operationalizing this mode of engagement; and
(3)~\textsc{Loom}, a creative writing probe in which authors orchestrate simulated narrative agents.

%% ============================================================
%% AI as Active Creative Medium
%% ============================================================

\section{AI as Generative Medium}
 
What does it mean to treat generative AI as an active medium for creativity? We use \textit{medium} to denote a mode of creative practice and \textit{material} to denote the substance being worked: generative AI is the material, and orchestrating it constitutes the medium. The framing draws on Sch{\"o}n's account of professional practice~\cite{schon1983}. Sch{\"o}n argued that skilled practitioners do not solve problems by applying rules to well-defined inputs, but through an ongoing dialogue with their situation. An architect sketches a layout, reads what the sketch reveals, and revises; a therapist offers an interpretation, observes the client's response, and adjusts. In each case, the practitioner's expertise lies not in producing correct outputs but in sustaining a productive conversation with complex, responsive material.
 
The idea that AI capabilities function as a design material is not new. Yang et al.\ explored how NLP presents unique challenges as a design material, requiring new sketching methods to reason about language interactions at appropriate levels of abstraction~\cite{yang2019sketching}. Kulkarni et al.\ found empirically that text-to-image prompts serve as reflective design material, facilitating the kind of exploration and iteration that Sch{\"o}n described~\cite{kulkarni2023prompts}. Dalsgaard frames generative AI through pragmatist philosophy as providing ``instruments of inquiry'' that shape designers' perception and conception of a problem space, while warning that the fidelity of AI outputs risks fixating designers on surface refinement rather than problem framing~\cite{dalsgaard2025tft}. From cognitive science, Davis et al.\ proposed an enactive model of creativity grounded in the theory that creative cognition emerges through continuous, embodied interaction between agent and environment---not through planning followed by execution~\cite{davis2015enactive}. Their model foregrounds \textit{directives} (vague creative intentions refined through interaction) over goals (linear plans), and emphasizes participatory sense-making as the locus of creative collaboration.
 
Our framing builds on these foundations but addresses a challenge they did not face: the active convergence tendency of large language models. Where prior accounts of AI-as-material treat the material as essentially passive or neutral, generative AI actively drifts toward resolution. The medium framing therefore adds disruption and emergent recognition as first-class cognitive moves, operationalized in SOSS. Moreover, where Davis et al.'s enactive model positions human and computer as co-creative partners sharing the creative act, the medium approach maintains a clearer asymmetry: the human orchestrates the conditions for emergence while the AI simulates possibilities within those conditions. This preserves a form of authorial ownership that co-creative framings can obscure.
 
Sch{\"o}n distinguished two modes of reflective engagement: \textit{reflection-on-action}, where the practitioner steps back to evaluate after the fact, and \textit{reflection-in-action}, where thinking and doing are fused. The trust calibration approach relies almost entirely on reflection-on-action: the human pauses, evaluates AI output, and judges whether to accept or override. The medium approach involves both modes: the orchestrator steps back to reshape characters or rethink a scenario (reflection-on), but also reads and responds to the simulation in real time (reflection-in). The difference is not which mode is used, but \textit{what the reflection is directed at}: not evaluating output correctness, but shaping the conditions from which interesting possibilities emerge. This preserves human creative agency: the orchestrator determines the conditions under which interesting things emerge, rather than every instance the system produces, much as designers of generative art own the system rather than its every output~\cite{galanter2016}.
The SOSS framework we introduce below operationalizes this shift, adding disruption and emergent recognition as first-class cognitive moves alongside shaping and observation.
 
The tendency of large language models to generate toward convergence is precisely what makes the human role essential. LLMs are systematically sycophantic: they gravitate toward agreement, smooth over conflict, and resolve tension prematurely~\cite{sharma2024sycophancy}. Left to run, a naive simulation produces predictable, harmonious outcomes. Unlike inert craft materials, this medium actively drifts toward resolution. The human orchestrator's role is to stir the system: inject conflict, remove easy resolutions, introduce asymmetries, and recognize when something interesting has emerged. The material's tendency toward blandness is itself a productive resistance, creating the friction that sustains creative work.
 
This reframing also relocates the site of effort in human-AI interaction. Trust calibration asks the human to expend effort judging whether AI output is correct or reliable, a form of vigilance that scales poorly and transfers little to future tasks. The medium approach asks the human to struggle with shaping conditions, reading emergent behavior, and deciding when to intervene. Such efforts develop a practitioner's feel for the material. Where evaluation develops discernment about \textit{this output}, orchestration develops judgment about \textit{this kind of system}, rendering it a transferable, deepening skill.

%% ============================================================
%% Loom: A Creative Writing Probe
%% ============================================================

\section{\textsc{Loom}: A Creative Writing Probe}

We use creative writing as a probe domain for two reasons.
First, it is a creation task; the category where Vaccaro et al.\ found human-AI synergy gains~\cite{vaccaro2024}. 
Second, narrative simulation (agents with goals, beliefs, and relationships interacting over time) provides a natural foundation for medium framing, as demonstrated by work on generative agents~\cite{park2023generative}.

\begin{figure}[tbh]
    \centering
    \includegraphics[width=0.9\columnwidth]{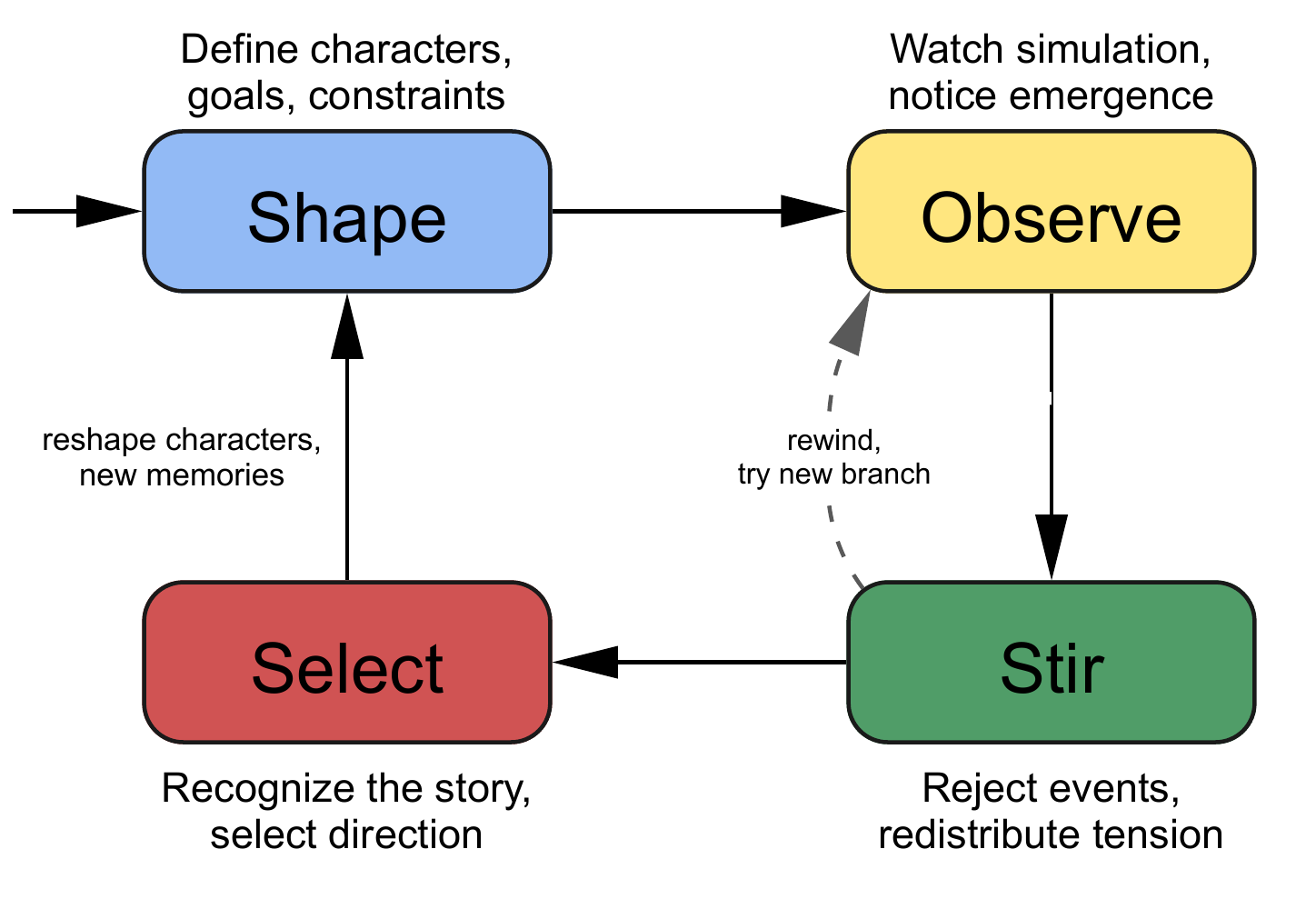}
    \caption{\textbf{The SOSS framework as an iterative cycle.}
        Authors enter through \textbf{Shape} by defining initial conditions---characters, goals, and constraints---then cycle through \textbf{Observe}, \textbf{Stir}, and \textbf{Select}.
        Selection reshapes conditions for subsequent cycles.
        The dashed arrow indicates a tight inner loop: authors frequently alternate between observing and stirring before committing to a selection.}
    \label{fig:soss}
\end{figure}

\begin{figure*}[t]
    \centering
    \includegraphics[width=1\textwidth]{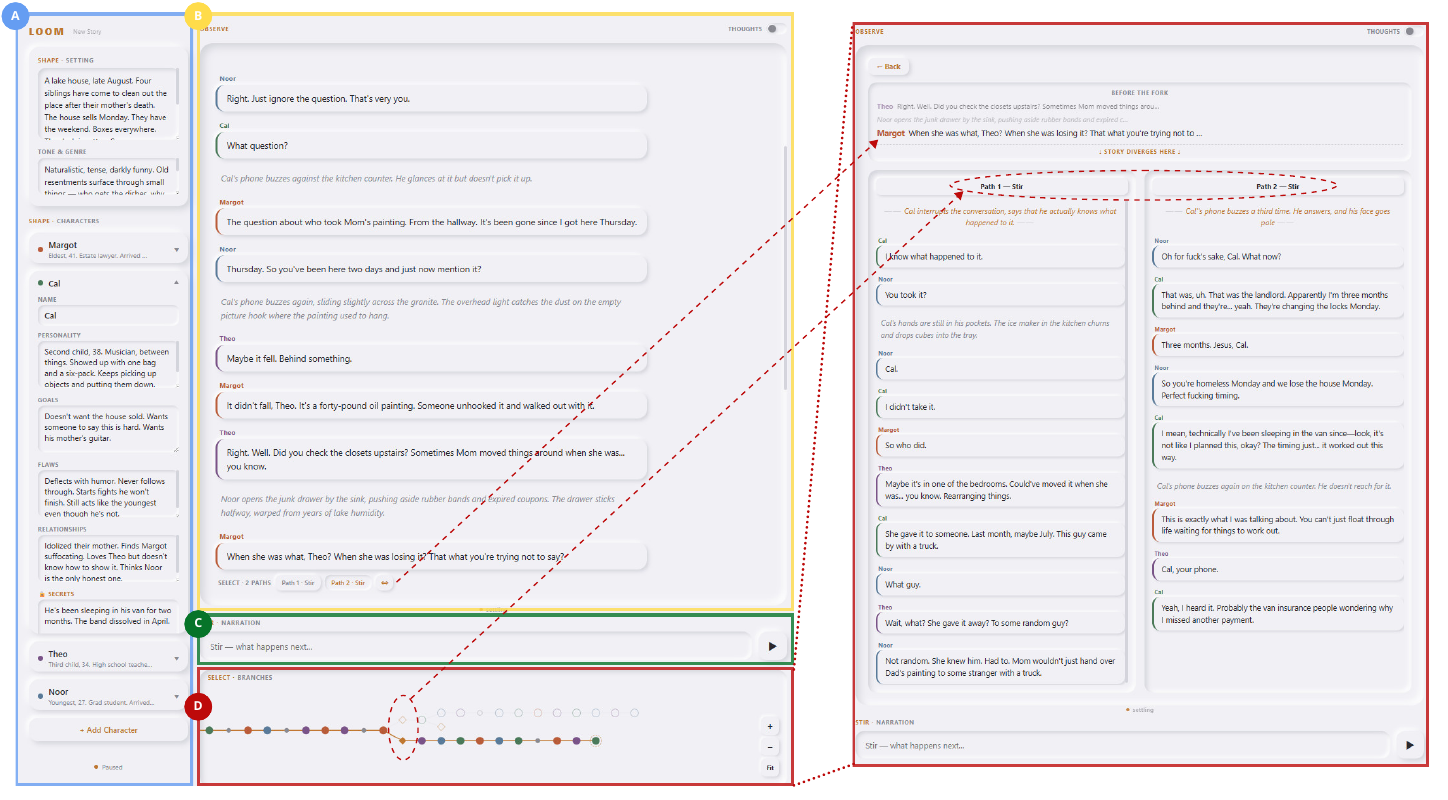}
    \caption{\textbf{The \textsc{Loom} interface.}
    \textbf{(A)~Shape:} authors define the setting, tone \& genre, and character profiles (shown expanded for Cal, with personality, goals, flaws, relationships, and secrets).
    \textbf{(B)~Observe:} characters (LLM agents) converse autonomously while italicized stage directions narrate actions and scene details; a path selector at the bottom of the panel lets the author switch between alternate branches.
    \textbf{(C)~Stir:} a text bar for injecting stir events into the simulation at any point.
    \textbf{(D)~Select:} a branch timeline where each node represents a story turn; the expanded view (right) replays the shared context before the fork and displays two divergent paths side by side, each produced by a different stir intervention, allowing the author to compare outcomes and choose which direction to develop.}
\label{fig:loom}
\end{figure*}

We present \textsc{Loom}, a system that makes the active medium framing concrete as a creative writing interface.
The interaction in Loom follows four cognitive phases, which we term the SOSS framework as shown in Figure~\ref{fig:soss}.
The first two phases, \emph{Shape} and \emph{Observe}, draw on Sch\"{o}n's reflective conversation~\cite{schon1983}, defining the situation and reading the material's response.
The latter two, \emph{Stir} and \emph{Select}, address challenges specific to generative AI as an active medium: disrupting the system's tendency toward convergence, and recognizing emergent direction worth pursuing.

\textbf{Shape.} The author defines the conditions for emergence: who the characters are, what situation they inhabit, and what constraints apply.
The author can pause the simulation at any point to revise a character's goals, add a flaw, or restructure relationships.

\textbf{Observe.} The author watches the simulation unfold.
Each character operates as an independent LLM agent, speaking based on their identity and what they have perceived, producing exchanges the author did not script.
The author's task is to read the material and notice what is interesting.

\textbf{Stir.}
The author intervenes when the simulation settles into predictable patterns, resolves tension too easily, or simply when the author has an idea.
Stirring happens through the narrator bar: the author types events that enter the shared world (``A letter falls from Erik's coat,'' ``The power goes out,'' ``Three hours pass in silence'').
These are stage directions that all characters perceive and possibly react to.
Because LLM-based simulations tend toward convergence, stirring is not occasional maintenance but a core authorial activity.

\textbf{Select.} Selection is based on the author's recognition: the accumulated decision to continue developing one narrative direction rather than resetting or diverging.
The branch timeline, along with buttons to compare branches (Figure~\ref{fig:loom}(D)), supports this by allowing authors to compare divergent paths side by side before committing.

In a typical session (Figure~\ref{fig:loom}), an author defines a cast of characters with conflicting goals and hidden information, then sets a scenario in motion. If characters resolve their tension too quickly---the inherent convergence tendency of LLMs---the author can stir, introducing an event that reopens conflict, and branch the timeline to compare how the scene unfolds under different conditions. From there, the author can select a direction, reshape characters to reflect new pressures, and continue.

Memory is selective; characters collect memories from interactions and events.
Recent exchanges persist in a working memory window of five messages, while older messages are evaluated for emotional significance using an LLM-rated impact score.
Messages exceeding a threshold are retained as long-term memories, potentially altering their understanding of another character; the rest fade.
This produces a simplified yet naturalistic pattern of forgetting and fixation that the author can explore through narration.
``Thoughts'' in the top-right corner of Loom can be enabled to display these memories and their underlying thoughts to the author.

\section{Discussion}

The medium framing raises several questions for the Tools for Thought (TfT) workshop and human-AI interaction. We discuss three here, aimed at opening avenues for future work rather than closing them.

\textbf{Strategy.}
If trust calibration consistently underperforms in decision tasks while creation tasks show gains~\cite{vaccaro2024}, should we be designing more systems around creation-style interaction?
Recent empirical work suggests yes: Reicherts et al.\ found that AI which extends users' own reasoning integrated better into decision-making and led to more reflective, higher-quality outcomes than AI which provides direct recommendations~\cite{reicherts2025ai}.
The medium framing constitutes a form of such process-oriented support, and the design principles that follow are concrete: give users control over \emph{conditions} rather than outputs; make the material's convergence tendencies visible and disruptable; and treat friction as a design goal rather than a failure mode.
This echoes emerging arguments for AI-in-the-loop systems where the human remains the primary agent~\cite{natarajan2024humanintheloopaiintheloopautomatecollaborate}, treating AI as a way to empower users rather than replace them~\cite{shneiderman2022hcai}.

\textbf{Outcomes.} When the human role shifts from judge to orchestrator, evaluation must follow.
Standard metrics for human-AI collaboration, such as task accuracy, decision quality, and calibration error, measure discernment about individual outputs.
But orchestration develops judgment about a kind of system: which configurations produce tension, when to intervene, what opens possibilities versus closing them.
This is a transferable, deepening skill.
Relating this to the TfT outcome framework~\cite{tankelevitch2025understanding}, the medium framing produces all three outcome types, but inverts their usual priority.
The simulation branches and character configurations the author creates in Loom are \emph{intermediary outcomes}: artifacts that externalize exploration and scaffold thinking.
The orchestration skill that develops over repeated sessions is a \emph{cognitive outcome}.
And the final narrative is a \emph{task outcome}, but notably the least important of the three.
In the medium approach, the outcome of using a TfT is less the artifact produced (the story, in Loom's case) than the way interaction with the material reshapes the author's creative process.
We suggest that evaluation should attend to the development of orchestration expertise over time and what we call \emph{felt understanding}, the metacognitive sense that one's engagement with a creative problem was generative, providing a deeper grasp of the problem space rather than mere acceptance of a good-enough AI output.
The orchestrator must monitor their own creativity, noticing when they are settling for convenience, when they are genuinely surprised, and when to push further.
This form of metacognition could serve as a metric for evaluating medium-style interaction tasks.
Process tracing and think-aloud protocols may reveal emergence recognition; longitudinal designs could track skill development. 

\textbf{Friction and adoption.} A recurring concern in human-AI interaction research is that cognitively beneficial friction meets user resistance~\cite{bucinca2021, ashktorab2025emergingreliancebehaviorshumanai}.
The medium approach reframes this tension.
In most generative AI tool designs, friction is an added cost imposed on the user for pedagogical or epistemic reasons, a cognitive forcing function that interrupts a workflow the user would prefer to be smooth.
In the medium approach, friction is intrinsic to the material.
LLMs tend toward convergence; the author encounters this not as an imposed interruption but as a property of the medium they are working with, much as a potter encounters the clay's tendency to collapse.
The need to stir---to inject conflict, break easy resolutions, sustain tension---arises from the work itself.
This may also address a broader concern raised by Lee et al.\, showing that generative AI use reduced self-reported cognitive effort and critical thinking among knowledge workers~\cite{lee10.1145/3706598.3713778}.
In the medium approach, the cognitive contribution is structural: orchestrators shape conditions rather than edit outputs, so the effort they experience reflects active creative engagement rather than passive consumption or judgment of final output.
This suggests that one strategy for incorporating metacognition into human-AI interaction may be to design systems in which cognitive engagement arises from the structure of the task, rather than from evaluating finished AI outputs.
This builds on deliberate reflective friction strategies~\cite{dalsgaard2025reflective, sarkar10.1145/3649404}, while suggesting that the most durable friction may be that which is intrinsic to the medium rather than imposed on the workflow. 

Although we used narrative orchestration as a probe, the medium framing is not limited to fiction.
Any domain where the goal is to explore a possibility space rather than produce a single correct output is a candidate. 
A clarifying diagnostic for any human-AI creative system may be to ask, following Sch\"{o}n~\cite{schon1983}: does this interface demand reflection-in-action or reflection-on-action?
If the latter, the system may be locating human effort in the wrong place.

\bibliographystyle{ACM-Reference-Format}
\bibliography{references}

\end{document}